# Rotation curves velocities obtained by warm low-density plasma simulating dark halos and 'dark matter'.

Y. Ben-Aryeh

Technion-Israel Institute of Technology, Physics Department, Israel, Haifa 32000

E-mail: phr65yb@physics.techion.ac.il ; orchid: 0000-0002-6702-2530

The free electron model with Boltzmann statistics for spherical low-density plasmas is developed further with asymptotic relations obtaining the density of electrons, mass densities, and the potentials of such plasmas. Solutions are developed as function of a pure number $x$ proportional to the distance from the stellar plasma center (galaxy center) with extremely small coefficient, so that these solutions are essentially functions of large astronomical distances and masses. The present plasma is divided into a central part and very long tail, where the central part of the plasma shows an exponential dependence on the distance from the galaxy center, but a part of the large mass of this plasma is included in the long stellar plasma tail. The present model is specialized to completely ionized Hydrogen plasma (with a small correction factor considering its mixture with heavier atoms) where emission and absorption of spectral lines can be neglected in the warm low density stellar plasma. We apply the present approach for treating rotation curves measurements, A general theory for rotation curves should include the superposition of the gravitational potentials introduced by the high-density compact stars, with those of the low-density stellar plasma potentials. But for halos which are at extremely large distance from the galaxy center, the dominant effects would be those of dark matter, and such dark halos are permeating and surrounding the compact galactic stars. Such plasma is found to be transparent in most of the EM spectrum. The existence of a large mass for the warm low-density plasma may solve the problem of "missing mass" in rotation curves measurements.

**Keywords**

## 1. Introduction

Rotation curves (RC) describe the rotational velocities of objects in a galaxy as function of their radial distance from the galaxy center. Various methods for deriving RC were described in the literature on this subject [1-3]. By studying many galaxies. it was found that the RC velocities are nearly constant, or "flat" with increasing distance away from the galactic center. This result is counterintuitive since based on Newtonian motion the rotational velocities would decrease for distances far away from the galactic center. By this argument the flat rotation curves suggest that each galaxy is surrounded by large amount of dark matter. There are various models describing dark matter haloes [4-10]. In the previous works [11, 12] I related the dark matter and the dark halos to warm low density ionized transparent plasma. The absorption and emission of such plasma is proportional to products of the electrons and ions extremely low densities so that such effects can be neglected. But as such low-density plasma extends over extremely large volumes their nucleons masses are significant, and their gravitational effects become important.

The physics of the high-density stars is well understood. Energy generation is in the hearth of stars. It provides the energy that we see as light, and it also supplies the heat and pressure that supports stars structures. The power source for stars is the thermonuclear fusion which leads to extremely high temperature in the cores of such stars. On the other hand, the source of energy in dark matter is not clear. Much effort was spent both theoretically and experimentally to find new kinds of non-baryonic interactions leading to dark matter, but there is not yet any evidence to the existence of such forces. In my previous work [12], I related the production of the warm low-density plasma to interaction between the low density interstellar gas and cosmic rays.

While the high-density stars exist in relatively small parts of the galaxy, the dark matter exists everywhere in the galaxy, and beyond it as dark halo. In the present analysis, this dark matter is represented by very low-density warm stellar plasma which is invisible but has large gravitational forces due to its extreme volume. In the present analysis it has been claimed that the total attractive force affecting the rotation curves measurements is produced by the superposition of the Kepler potential, and that of the low-density stellar plasma potential.

A general theory of RC velocities measurement should include the superposition of the gravitational forces produced by the high-density compact stars, with that of the dark halo. In the present work I restrict the analysis to non-rotating dark halo. For halos which are at extremely large distances from the galaxy center, and which are beyond the compact stars the dominant effects will be mainly those of the dark halos. We consider the conclusion in astronomy, that the dark halo mass is equal approximately to 4 times the Kepler mass. We concentrate in the following analysis on the effects of the dark halos at very long radial distances in which the Kepler forces are quite small, but they should be taken into account in the general analysis.

The present paper is arranged as follows: In Section 2, the stability of warm low density transparent Hydrogen plasma is analyzed by Boltzmann statistics, including asymptotic relations. The present analysis has the advantage, that it is developed from first physical principles, while other descriptions of the halo density profiles [4-10], simulate gravitational observations without information about the composition of this matter. In Section 3, we analyze the Rotation curves phenomena and relate them to the Halo mass produced by the warm low density Hydrogen plasma. In Section 4 we summarize our results with a conclusion.

## 2. The stability of warm low-density transparent Hydrogen plasma including the Halo surrounding our galaxy, analyzed by Boltzmann statistics with asymptotic relations

Following the explanations given in the introduction a Halo is surrounding our galaxy, composed mainly from warm low-density completely ionized Hydrogen plasma. For simplicity of discussion, we treat the Halo low-density stellar plasma with spherical symmetry, where its stability is obtained by a balance between gravitational forces and plasma pressure.

The gravitational force $g$ per unit mass at a distance $r$ from the spherical stellar plasma center is due entirely to the mass $M_r$ interior to this distance:

$$g = -GM_r / r^2 \tag{1}$$

where $G$ is constant of gravitation. We denote $r$ and $\phi$ as the distance and the potential of this plasma from its center, respectively. Assuming that for this stellar plasma, gravitational potential $\phi$ has spherical symmetry, then:

$$g = -d\phi(r)/dr \quad . \tag{2}$$

According to the hydrostatic equation for isothermal free electron model

$$dP = g\rho dr \quad . \tag{3}$$

where $P$ is the stellar plasma pressure and $\rho$ the stellar-mass density, both are functions of distance $r$ from the stellar plasma center (galaxy center). This equation describes the decrease of the plasma pressure (proportional to decrease of the averaged ionized electron density) as we move to larger values of $r$ opposing the attractive gravitational forces.

Under the condition that the stellar plasma behaves as an ideal gas, then the pressure $P$ is given by $P = n_i k_B T$. Assuming also that the gradient of temperature is small relative to the gradient of the electron density $n_e$ (isothermal process) then we get:

$$dP = k_B T dn_i = -\rho d\phi \qquad . \qquad (4)$$

The density of the plasma is obtained by following the semi-empirical relation:

$$\rho(r) = n_i(r) \kappa m_N \qquad (5)$$

where $n_i(r) (m^{-3})$ and $\rho(r) (kg/m^3)$ are, respectively, the ionized electrons density and the mass density given as function of the distance $r$ from the stellar plasma center, $\kappa \approx 1.5$ is a correction term which considers the mixture of Hydrogen atoms with Helium and heavier atoms [13], and $m_N = 1.67 \cdot 10^{-27} kg$ is the Hydrogen nucleon mass.

By substituting Eq. (5) into Eq. (4). We get:

$$k_B T dn_i(r) = -\kappa m_N n_i(r) d\phi(r) \rightarrow \frac{dn_i(r)}{n_i(r)} = -\frac{\kappa m_N}{k_B T} d\phi(r) \qquad . \qquad (6)$$

This equation has the general solution:

$$n_e(r) = n_0 \exp\left[-\left(\frac{\kappa m_N}{k_B T}\right)\phi(r)\right] \quad ; \quad \phi(r=0) = 0 \quad ; \quad n_i(r=0) = n_0 , \qquad (7)$$

where $n_0 (m^{-3})$ is the ionized electrons density in the center of the stellar plasma (taken as experimental parameter), and by our definition the potential $\phi$ vanish at the stellar plasma center. We choose the potential to vanish at the center of the galaxy and not at infinity in which the conditions are not well defined.

One should notice that the potential $\phi(r)$ is given in unit of $\left(\dfrac{k_B T}{\kappa m_N}\right)$ i.e. unit of velocity squared ( $m^2/\sec^2$ ) as the exponent in Eq. (7) is a pure number. The potential $\phi$, for a Halo with spherical symmetry, satisfies the Poisson equation:

$$\frac{d^2\phi(r)}{dr^2} + \frac{2}{r}\frac{d\phi(r)}{dr} = -4\pi G\rho(r) \quad . \tag{8}$$

By substituting Eqs. (5) and (7) into Eq. (8) weget:

$$\frac{d^2\phi(r)}{dr^2} + \frac{2}{r}\frac{d\phi(r)}{dr} = 4\pi G\kappa m_N n_0 Exp\left[-\left(\frac{\kappa m_N}{k_B T}\right)\phi(r)\right]. \tag{9}$$

Substituting in Eq. (9):

$$x = r\sqrt{\xi} \quad ; \quad \xi = 4\pi G\frac{(\kappa m_N)^2 n_0}{k_B T} = 1.69\cdot 10^{-40}\frac{n_0\kappa^2}{T} \quad (m^{-2}) \tag{10}$$

we get:

$$\frac{d^2\phi(x)}{dx^2} + \frac{2}{x}\frac{d\phi(x)}{dx} = \left(\frac{k_B T}{\kappa m_N}\right)Exp\left[-\left(\frac{\kappa m_N}{k_b T}\right)\phi(x)\right]. \tag{11}$$

One should take into account that $x$ is a normalized pure number, where for $x=1$ the distance $r$ from the plasma center stretches to extremely very long distance equal to: $1/\sqrt{\xi}$ , and where according to Eq. (10) $\sqrt{\xi}$ is a very small number. Our solutions for the number of ionized electrons and the potentials, are given as function of the pure numerical parameter $x$ i.e. as $n_e(x)$ , and $\phi(x)$. So, by transforming our solutions to be functions of the distance $r$ from the plasma center, they become functions of astronomical long distances.

In the previous work [11], by using Eq. (7), we performed the first and second derivatives of $n_i(r)$ according to $r$ and related such derivatives to the corresponding derivatives of $\phi(r)$. By substituting these relations in Eq. (11), and using the relation $x = r\sqrt{\xi}$ , we obtained the differential equation for $\theta(x) = \dfrac{n_i(x)}{n_0}$ which is given as [11]:

$$\frac{\partial^2 \theta(x)}{\partial x^2} - \frac{2}{x}\frac{\partial \theta(x)}{\partial x} + \frac{1}{\theta(x)}\left(\frac{\partial \theta(x)}{\partial x}\right)^2 = \theta(x)^2 \quad ; \quad x = r\sqrt{\xi} \quad ; \quad \theta(x) = \frac{n_i(x)}{n_0} \quad . \quad (12)$$

It is quite easy to find that Eq. (12) is satisfied for

$$\theta(x) = \frac{n_i(x)}{n_0} = \frac{2}{x^2} \rightarrow n_i(x) = n_0 \frac{2}{x^2} \quad ; \quad n_i(r) = n_0 \frac{2}{\xi r^2} \quad . \quad (13)$$

Here the parameter $x = r\sqrt{\xi} \gg 1$ where $r$ is the distance from the plasma center (galaxy center) and $\xi$ is avery small number.  But as the solution of Eq. (12) by Eq. (13) does not satisfy the boundary equation $\theta(x=0) = \frac{n_i(x=0=r)}{n_0} = 1$ , this solution is valid only for $x \gg 1$.

An approximate solution is obtained for the region near the center of the star by using series expansion of the exponential function of Eq. (11).  Then, we get:

$$\theta(x) = \frac{n_e(x)}{n_0} = Exp\left[-\frac{1}{8}x^2\right] \rightarrow \frac{n_e(r)}{n_0} = Exp\left[-\frac{1}{8}\xi r^2\right] \quad ; \quad \xi \simeq 1.69 \cdot 10^{-40} \frac{\kappa^2 n_0}{T} \quad (m^{-2}) \quad . \quad (14)$$

These solutions enable us to get asymptotic results at large distance $r$ from the star center, for the density of ionized electrons, as given by Eq. (13). In addition to the central part of such stellar plasma given by Eq. (14), we get a very long tail of the stellar plasma potential and its density as given by Eq. (13),

By using numerical calculations of Eq. (12) we find that the approximation (14) is valid near the stellar plasma center (for $x \leq 4.07$ ). The numerical calculations for $x \geq 4.07$ were found to be equal approximately to the solutions by Eq. (13). We can combine these asymptotic solutions into an approximate analytical result for all the range of $x$ values as:

$$\theta(x) = \frac{n_i(x)}{n_0} = Exp\left[-\frac{1}{8}x^2\right](for \quad 0 \leq x \leq 4.07) + \frac{2}{x^2}(for \quad 4.07 \leq x \leq \infty) \quad . \quad (15)$$

The use of the function: $\theta(x) = \frac{n_e(x)}{n_0}$ is remarkable. It gives also the plasma mass density $\rho(kg \cdot m^{-3}) = n_e \kappa m_n$ as function of the distance $r\ (m) = \frac{x}{\sqrt{\xi}}$ from the stellar plasma center, which is proportional to the electron density $n_0(m^{-3})$ at its center. One should notice that the density of electrons $n_i(x)$ for $x \gg 1$ is decreasing inversely proportional to $x^2$. We calculated the mass of the stellar plasma separately for its central part and for its long tail.

**The number of electrons in the central part of the low-density star plasma**

Using Eq. (14) the total number of electrons $N_{e,cent}$ in the central part of the present plasma is calculated as:

$$N_{e,cent} = n_0 \int_0^{4.07} Exp\left[-\frac{1}{8}x^2\right] 4\pi r^2 dr = \left(4\pi n_0 / \xi^{3/2}\right) \int_0^{4.07} \exp\left[-\frac{1}{8}x^2\right] x^2 dx \quad . \tag{16}$$

**The number of electrons in the tail of the low-density star plasma**

According to Eq. (15), the number of electrons in the tail of the low-density star plasma $N_{e,tail}$ is given by

$$N_{e,tail} = n_0 \int_{R_{min}}^{R_{max}} \frac{2}{\xi r^2} 4\pi r^2 dr = \frac{8\pi n_0 (R_{max} - R_{min})}{\xi} \quad ; \quad R_{min} = \frac{x_{min}}{\sqrt{\xi}} = \frac{4.07}{\sqrt{\xi}} \quad . \tag{17}$$

We find that under the limit $R_{max} \to \infty$ the number of electrons in the tail of the low-density plasma is diverging. In rotation curves measurements one considers the number of electrons that are interior to the point of measurements so that $R_{max}$ is the distance from the point of measurement to the center of the galaxy.

## 3. Rotation curves related to warm low density Hydrogen plasma

Considering a small mass with mass $m(kg)$ rotating in circular motion with radius $R$ around a big spherical mass $M(kg)$ then the Kepler potential is given by $m\frac{GM_{R,stars}}{R}$ where $M_{R,stars}$ is the big spherical mass within a distance $R$. The rotational kinetic energy of the small mass is given by $\frac{1}{2}mv_{rot}^2$. Due to the virial theorem for Kepler motion the kinetic energy is equal one half of the potential energy so we get

$$\frac{1}{2}mv_{rot}^2 = \frac{1}{2}m\frac{GM_{R,stars}}{R} \to v_{rot}^2 = \frac{GM_{R,stars}}{R} \quad . \tag{18}$$

Eq. (18) gives the Kepler rotational velocity under the approximation of neglecting deviation from spherical symmetry. By adding to the galaxy stars mass $M_{R,stars}$ the dark matter halo $M_{R,halo}$ (described as dark matter) Eq. (18) should be exchanged into

$$\frac{1}{2}mv_{rot}^2 = \frac{1}{2}m\frac{GM_{R,stars}}{R} + \frac{1}{2}m\frac{GM_{R,halo}}{R} \rightarrow v_{rot}^2 = \frac{GM_{R,stars} + GM_{R,halo}}{R} \quad . \tag{19}$$

Since the galaxy stars have usually a disk shape the use of Eq. (19) can be considered only as rough approximation. On the other hand, we estimate that the dark halo has spherical symmetry as might follow from gravitational profiled related to dark haloes [4-10]. Due to the existence of dark matter the gravitational forces operating on the mass $m(kg)$ are composed of two parts: a) The Kepler attractive forces between the small mass and the galaxy high density stars. b) There is an additional attractive force between the small mass and the invisible dark matter mass (according to the present analysis low-density plasma) included within a distance R. While the idea that RC velocity measurements are affected by such dark matter is quite common in the published literature [1-10], the physical nature of such matter is not clear and is, today, under many debates. One should consider that historically the behavior of RC measurements which is in contradiction with Kepler motion was the first reason for introducing the dark matter, which was considered as the missing mass (see e.g. [14]).

It is interesting to find that the mass density profile of the present work given by Eq. (15) is similar to that obtained by Burkert [7] which studied the structure of dark matter halos in dwarf galaxies and arrived at isothermal density profile:

$$\rho(r) = \frac{\rho_0}{1 + (r^2/r_e^2)} \quad , \tag{20}$$

where $r_e$ is the core radius and $\rho_0$ is the density in the center of $\rho(r)$. This density profile is similar to the present low density profile as the tail of Eq. (13) proportional to $1/x^2$ can be related to Eq, (20) by the equality $1/x^2 = 1/(r\sqrt{\xi})^2 = \frac{1/(\sqrt{\xi})^2}{r^2} = \frac{(r_0)^2}{r^2}$ where $1/\sqrt{\xi} = r_0$ represents the central stellar plasma radius. Also, in the center of low-density stellar plasma we have density $n_0 \kappa m_N$ which is given in Eq. (20) by $\rho_0$ but the present density profile has a certain exponential dependence near the star plasma center (see Eq. (16)). Burkert in the abstract of his article [7] claimed: "that dark matter must be baryonic." We should take into account that for the total mass $M_{halo}$ of the halo of Eq. (20) we need to use the integral $M_{halo} = 4\pi \int_0^\infty r^2 \rho(r) dr = \int_0^\infty \frac{4\pi r^2 \rho_0}{1+(r^2/r_e^2)} dr$ and this integral will slowly diverge for $r \to \infty$. Some authors eliminated this divergence by adding high powers of $r$ in the denominator of Eq. (20). This divergence is prevented in the present model of low-density plasma as $R_{max}$ is fixed by the distance of the measurement point to the center of the galaxy and very high temperatures are eliminated by the invalidity of the Boltzmann distribution.

## Numerical calculations of the halo mass produced by warm low-density plasma.

### A. For the central part of the low-density plasma

The number of electrons in the central part region is given by performing the integration of Eq. (16) and substituting in this equation the value of $\xi$ according to Eq. (10). Then we get:

$$N_{e,cent} = (4\pi n_0) \cdot 7.554 \cdot \left(1.69 \cdot 10^{-40} \frac{\kappa^2 n_0}{T}\right)^{-3/2}$$
$$= 43.21 \cdot 10^{60} \left(\frac{T}{\kappa^2}\right)^{3/2} \frac{1}{\sqrt{n_0}} \quad (21)$$

The product of Eq. (21) with the Hydrogen nucleon mas of $1.67 \cdot 10^{-27} kg$, including the correction factor $\kappa$, gives the mass of the central part of the low-density plasma as:

$$M = 7.22 \cdot 10^{34} \frac{T^{3/2}}{\kappa^2} \frac{1}{\sqrt{n_0}} (kg) \quad (22)$$

By inserting typical values for the warm low-density plasma as: $T = 10^5 (K)$ ; $n_0 = 10^5 (m^{-3})$ ; $\kappa = 1.5$. we get:

$$M = 7.22 \cdot 10^{34} \frac{10^{5 \cdot 3/2}}{1.5^2} \frac{1}{10^2 \sqrt{5}} (kg) = 4.3 \cdot 10^{39} (kg) \quad (23)$$

which is equal to $2.15 \cdot 10^9$ the sun mass, approximately. We find the interesting result that the number of electrons as obtained from Eq. (21) is *inversely* proportional to $\sqrt{n_0}$. This result is related to the volume which is proportional to: $r^3 \propto \left(\sqrt{n_0}\right)^{-3}$, so that by multiplying this factor by the number of electrons $n_0$ we get the proportionality constant, for the number of photons (and correspondingly for the plasma mass), as $1/\sqrt{n_0}$. We should notice that for temperatures which are higher than those assumed in Eq, (23), and for lower density $n_0$, the mass $M$ for the central part of the plasma becomes larger.

### B. For the tail of the low-density plasma

For this case we integrate Eq. (17), from $r = R_{min}$ to $r = R_{nax}$, Then we get:

$$N_{e,tail} = n_0 \int_{R_{min}}^{R_{max}} \frac{2}{\xi r^2} 4\pi r^2 dr = 1.49 \cdot 10^{41} \frac{T}{\kappa^2}(R_{max} - R_{min}) ;$$

$$\xi \simeq 1.69 \cdot 10^{-40} \frac{\kappa^2 n_0}{T} \ (m^{-2}) ; \ R_{min} = \frac{4.07}{\sqrt{\xi}} = (3.13/\kappa) \cdot 10^{20} \sqrt{\frac{T}{n_0}} \tag{24}$$

We have the problem of finding the value of $R_{max}$. For treating rotation curves we can estimate $R_{max}$ as the distance of the measurement point (i. e. the location of the rotation curve) from the galaxy center. Taking for example the rotational curve measurements made in [1], they were obtained in radii range from 4 to 122 kpc. For $122 kpc = 3.76 \cdot 10^{21} m$ we get:

$$N_{e,tail} = 1.49 \cdot 10^{41} \frac{T}{\kappa^2}\left(3.76 \cdot 10^{21} - 2.09 \cdot 10^{20} \sqrt{\frac{T}{n_0}}\right) ; \ . \tag{25}$$

Under the conditions $T = 10^5 (K)$ ; $n_0 = 10^5 (m^{-3})$ ; $\kappa = 1.5$, Eq. (25) is transformed to:

$$N_{e,tail} = 0.596 \cdot 10^{46} \left(3.76 \cdot 10^{21} - 2.087 \cdot 10^{20}\right) \tag{26}$$

We find that under these conditions $R_{min}$ is smaller by order of magnitude relative to $R_{max}$. By multiplying Eq, (26) by $\kappa m_N = \kappa \cdot 1.67 \cdot 10^{-27} kg$, we get for the mass of the plasma tail:

$$M_{tail} = \left(5.6 \cdot 10^{40} - 3.12 \cdot 10^{39}\right) kg \tag{27}$$

Under these conditions the plasma tail mass is larger by an order of magnitude relative to its central part. *By considering smaller distances of the rotation curves to the galaxy center we arrive at a certain critical distance in which only the central part contributes to the plasma mass. Qualitatively we notice that the* *rotation curves show a rapid rise, corresponding to the exponential region of the plasma, and after that to a slow rise corresponding to the plasma long tail.*

## 4. Summary, discussion and conclusion

In previous works on warm low-density plasma [11,12], I have shown that dark matter phenomena can be related to such plasma. In [11] the stability of low-density stellar plasma was analyzed for a star with a spherical symmetry in equilibrium between the gravitational attractive forces and the repulsive pressure forces of an ideal electron gas, where the analysis was developed by using Boltzmann statistics. The absorption and emission of radiation for extremely low-density plasmas was found to vanish over the electro-magnetic (EM) spectrum, This work is supported by a recent article [12] in which it was shown that the interstellar gas is heated by the cosmic rays, and the warm ionized plasma is observed only by pulsar dispersion in the far radio frequency region. While for relatively low temperatures the state of molecular, atomic, and partly ionized plasma, are analyzed by the ISM model [13], with extensive EM spectra, we show that for higher temperatures and lower densities the plasma becomes transparent. This conclusion was obtained in agreement with results from conventional theories [14-17]. One of the main proofs for the existence of dark matter is obtained from rotation curves measurement, as explained in the introduction. The aim of the present work was to show that rotation curves measurements can be related to warm low-density plasma and the present analysis can be summarized as follows.

We extended the analysis of low density spherical stellar plasma [11] by which the density of ionized electrons $n_e (m^{-3})$, the mass density $\rho(r) (kg/m^3)$, and potential $\phi(r)$ are described as functions of the distance $r$ from the galaxy center. Using stability analysis with Boltzmann statistics the density of ionized electron $n_i(r)$ was found to have an exponential dependence on the gravitational potential $\phi(r)$ given by Eq. (7). Poisson gravitational equation for the potential $\phi(r)$ was given in Eq. (8). Differential equation was developed for the potential $\phi(x)$ in Eq. (11), where we exchanged the distance $r$ into a pure number parameter $x$. and where $x = r\sqrt{\xi}$, and $\xi$ is an extremely small number given by Eq. (10). Differential equation was obtained in Eq. (12) for the ionized electrons density $n_i(x)$ as function of $x = r\sqrt{\xi}$, and the density of ionized electrons $n_0$ in the stellar plasma center, taken as empirical parameter. The mass density of the central plasma profile was calculated by the product of the exponential equation

(14) for the density of electrons with the proton mass, for warm Hydrogen completely ionized plasma (with a small correction factor for its mixture with heavier atoms). The mass of the long Hydrogen stellar plasma tail was obtained by the product of the electrons density of Eq. (13) with the proton mass. We used here the asymptotic solution $n_i(x) = \frac{2}{x^2}$ for $x \gg 1$. The total mass in the central part of the Hydrogen plasma and its tail were obtained by integrations over the distances for which such plasma models are valid. The divergence obtained for the stellar plasma tail, slow dependence on $r$, was eliminated by choosing the upper bound of integration as the distance $R_{max}$ of the rotation curve from the galaxy center. For extremely long radial distances which are larger than this limit, we expect breakdown of the Boltzmann approximations, possibly due to escape of electrons from the interaction region (see e, g, in [18]). In conclusion from the present analysis, I find that *the total mass which is obtained for the warm low-density plasma can solve the problem of "missing mass" in rotation curves measurements.*

In astronomy it has been shown that most of the matter in the universe is in the form of dark matter which does not interact with the EM radiation. The physical composition of dark matter is not clear and there are many debates about it. In the present work it has been shown that low density plasma can enter into semi-equilibrium Boltzmann state (within astronomical time scales) and has the properties of dark matter. A general theory should include the superposition of the gravitational potentials introduced by the high density compact stars with those of the low density stellar plasma potentials. Such theories might be very complicated but for halos which are at extremely large distance from the galaxy center the dominant effects will be those of dark matter, and such dark halos were analyzed in the present work by relating them to low density stellar plasma. By using free electrons model it was shown in the present article and in the previous ones [11,12] that the low-density plasma is transparent, in most of the EM spectrum and can be considered as dark matter.

## Acknowledgement

The present study was supported by Technion-Mossad under grant No. 200716## Declaration

The author declares no conflict of interest.